\documentclass[12pt]{JHEP} 
\usepackage{epsfig}
\newcommand\fverb{\setbox\pippobox=\hbox\bgroup\verb}
\newcommand\fverbdo{\egroup\medskip\noindent%
			\fbox{\unhbox\pippobox}\ }
\newcommand\fverbit{\egroup\item[\fbox{\unhbox\pippobox}]}
\newbox\pippobox

\def\be{\begin{equation}}
\def\ee{\end{equation}}
\def\ba{\begin{array}{l}}
\def\ea{\end{array}}
\def\bea{\begin{eqnarray}}
\def\eea{\end{eqnarray}}
\def\eq#1{(\ref{#1})}

\def\del{\partial}

\def\gap#1{\vspace{#1 ex}}
\def\tgap{\vspace{3ex}}

\def\myitem#1{\gap3\noindent{#1}\gap2}

\def\nn{\nonumber\\}

\def\ni{\noindent}

\def\ket#1{| #1 \rangle}
\def\bra#1{ \langle #1 |}
\def\mat#1#2#3{ \langle #1 | #2 | #3 \rangle}
 
\title{Rolling tachyon solution of two-dimensional string theory}
\author{Gautam Mandal$^a$  and Spenta R. Wadia$^{a,b,*}$\\
{$^a$\it Department of Theoretical Physics,\\
Tata Institute of Fundamental Research,\\ 
Homi Bhabha Road, Mumbai 400 005, India.\\
$^b$Theory Division, CERN\\ 
CH 1211, Geneva 23, Switzerland.\\
($^*$On leave from Tata Institute of Fundamental Research)}
\\~~\\
E-mail: \email{mandal,wadia@theory.tifr.res.in,Spenta.Wadia@cern.ch}}

\preprint{\hepth{0312192}\\
CERN-TH/2003-302\\
TIFR/TH/03-27}

\abstract{We consider a classical (string) field theory of $c=1$
matrix model which was developed earlier in hep-th/9207011 and subsequent 
papers. This is a noncommutative field theory where the noncommutativity
parameter is the string coupling $g_s$. We construct a classical solution 
of this field theory and show that it describes the complete time history of
the recently found rolling tachyon on an unstable D0 brane.}

\begin{document}
\section{Introduction} 
Recently there has been considerable interest in the c=1 matrix model
arising from the identification of c=1 matrix model \cite{V1,KMS} as a
non-perturbative description of open string dynamics on unstable D0
branes of two-dimensional string theory (see \cite{takayanagi,douglas}
for the fermionic version). The main merit of the matrix model
description is that it provides a (holographic) non-perturbative
formulation of two-dimensional string theory. The unstable D0 brane is
identified with a non-relativistic fermion (the perturbative
fluctuations correspond to relativistic fermions of the matrix model,
see below).

The main point of Klebanov, Maldacena and Seiberg
\cite{KMS}  is that it is necessary to treat the fermions
quantum mechanically ($\hbar = g_s=$ finite), in order to obtain
finite answers for quantities related to the decay of the D0 brane
into closed strings.  In this note we reinterpret the result of KMS in
terms of the classical solutions of a field theory that is exactly
equivalent to the c=1 matrix model. This field theory is
`noncommutative' because it takes into account the quantum mechanics
of the fermions. Most of the formalism and the time dependent
classical solution that we will discuss are known for some time
\cite{DMW-main,DMW-instanton}.  Here we will present an interpretation
of the solution as a rolling tachyon. Our formalism enables us to
write down the classical solution for all times.  There appears a
characteristic time scale $T_c = o(1)$ in the classical
solution. For $t \ll T_c$ the solution can be identified with a D0
brane. For $t \gg T_c$ the solution can be identified as a
perturbation of the filled Fermi sea which are directly mapped
\cite{DMW-beta} to closed string tachyon fluctuations. 
It is worth emphasizing that the `classical solutions' we are
discussing incorporates a dependence on the string coupling $g_s$,
because the field theory is non-commutative. They are different from
the classical solutions of the underlying fermions, which are
described by hyperbolas in a classical phase space (see section 2.).
We include a review of relevant parts of our earlier work on matrix
models in the Appendix.

In this paper we do not worry about the non-perturbative instability
of string theory described by the potential given in fig. 2. Our
discussion can be easily adapted to the case of a symmetric potential
(see \cite{DMW-discrete}, or for a fermionic interpretation,
\cite{takayanagi,douglas}).

We would like to mention that in \cite{gutperle} there was an attempt
to describe the rolling tachyon as a solution of a hybrid collective
field theory, where the D0 brane collective coordinate is treated
separately from the density waves near the Fermi level. In this
treatment it was not possible to obtain the classical solution,
representing the fermion density, for all times. It is not clear to us
whether collective field theory \cite{das-jevicki} can, in principle,
address this issue. This is because, in its present form, it does not
seem to include the source terms of the open strings. For a different
approach to noncommutativity in two dimensional string theory see
\cite{jevicki}. While this paper was being written we received
\cite{boer} which also discusses the rolling tachyon in the c=1 matrix
model.

The matrix model picture of the unstable D0 brane has been used by
\cite{sen} as an evidence for a new duality between open strings on
unstable D branes and certain sectors of a closed string theory.  In a
sense our formulation of the classical two-dimensional string theory
provides a description of both sides of the duality at two limits.

\begin{figure}[ht]
\vspace{0.5cm}
\hspace{-1in}
\centerline{
   {\epsfxsize=10cm
   \epsfysize=5cm
   \epsffile{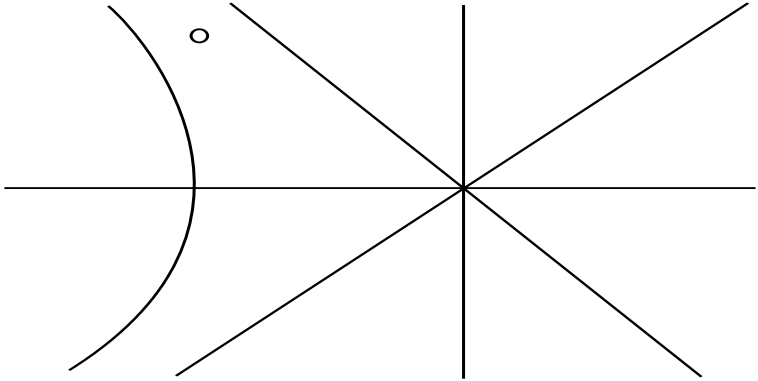}}
}
\caption{\sl  The classical solution $u_1$ represents a 
localized Gaussian solution in phase space far from the Fermi level. 
At late times, it can be identified with ripples
``near'' the Fermi surface; the process can be interpreted
as conversion into closed string modes.The support of the
Gaussian wave packet is non-zero because of uncertainty principle,
with $\hbar \sim g_s$, giving a finite decay amplitude.}
\label{phase}
\end{figure}

\begin{figure}[ht]
   \vspace{0.5cm}
\hspace{-1in}
\centerline{
   {\epsfxsize=10cm
   \epsfysize=5cm
   \epsffile{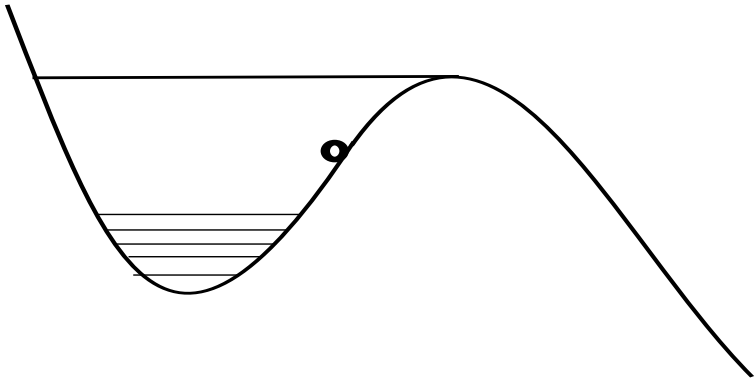}}
}
\caption{\sl Same phenomenon in coordinate space. The finite
localization in space implies finite width in energy.}
\label{phi3}
\end{figure}

\section{Classical Analysis }

We will first discuss the classical behaviour of the matrix model
which is related to the $g_s \to 0$ limit of the discussion in the
next section and also to the BCFT approach. In \cite{KMS,V1} the
following classical action for $N$ D0 branes has been introduced,
where $M_{ij}(t)$ describe the open string tachyon and $A_{0,ij}$
describes the gauge field: 
\be S= \frac1{g_0} \int dt\ {\rm Tr}[\frac12 (D_t M)^2 +
V(M)], \qquad D_t = \del_t - i [A_0, ]
\label{action}
\ee
\[ V(M) = -\frac12 M^2 + O(M^3)
\]
$1/g_0$ is the D0-brane mass. 
The classical equations of motion are
\[
\ddot M = M + O(M^2),
\quad
[M, \dot M] = 0
\]
The second equation is the Gauss law condition which 
reflects the gauge symmetry $M(t) \to U^\dagger(t) M(t) U(t),
A_0(t) \to U^\dagger(t) A_0(t)U(t) + i U^\dagger(t)\del_t U$.
In the following we will fix the unitary gauge:
\[
M(t) = {\rm diag}[q_i(t)]  
\]
If we ignore the $O(M^2)$ terms, the equation of motion becomes
\[
\ddot q_i = q_i
\]
with solution
\be
q_i(t) = q_i \cosh\ t + v_i \sinh\ t
\label{sinh}
\ee
$q_j, v_j$ can be interpreted as the initial position 
and velocity of the $j$-th D0 brane. 
Classically the D0 branes are non-interacting.

\myitem{\bf Hamiltonian:}

\ni Define the canonical momenta to be 
\[
p_i = v_i/g_0
\]
The Hamiltonian becomes
\[
H= \sum_i \frac12( g_0 p_i^2 - q_i^2/g_0)
= \frac1{g_0} \sum_i (\tilde p_i^2 - \tilde q_i^2)
\]
where $\tilde p_i = g_0 p_j, \tilde q_i = q_i$, so that
$\{\tilde p_i, \tilde q_i\} = \delta_{ij} g_0$.
Introducing the phase space density
\[
u(\tilde p, \tilde q)= \sum_i \delta(\tilde p -\tilde p_i)
\delta(\tilde q -\tilde q_i) 
\]
and the single particle Hamiltonian
\[
h(\tilde p, \tilde q)= \frac12 (\tilde p^2 - \tilde q^2)
\]
we can rewrite the Hamiltonian as
\[
H= \frac1{g_0} \int d\tilde p\ d\tilde q\ h(\tilde p, \tilde q)
u(\tilde p, \tilde q)
\]
The range of $\tilde p, \tilde q$ , in the above expressions, is the
entire plane. Hence each point in the ($\tilde p, \tilde q$) plane is a
possible classical state of the D0 brane of a given energy.
We will see below that, in the quantum theory, the D0 branes are interacting. 
This has a drastic effect on the spectrum of allowed states.

If the $O(M^3)$ terms are included in the potential, the hyperbolic
functions represent the initial behaviour. The $q_i$ will have an
oscillatory solution if it starts out inside the well on the left
side (see Fig. \ref{phi3}), and will reach infinity if it is on the
other side.

\subsection{Rolling tachyon}

The classical solution \eq{sinh} is interpreted by \cite{KMS,V1} as a
rolling tachyon. In \cite{KMS} the amplitude of such a configuration
to emit closed strings is calculated (a) using BCFT (combining earlier
rolling tachyon boundary state calculations of \cite{lambert,gaiotto}
and Liouville theory boundary state calculations in \cite{zamol}), and
independently (b) using matrix model (where the asymptotically valid
bosonization formulae for relativistic fermions were used to compare
with BCFT). It is found that the expectation value of the total
emitted energy diverges in the BCFT, whereas it appears to give a
finite answer in the matrix model. In the next section we will discuss
this in detail.

\section{The Fermion Field Theory or the $u(p,q,t)$ theory}

The action \eq{action} introduced above corresponds to the dynamics of
the singlet sector of the $c=1$ matrix model. As we reviewed in
Appendix A, the classical analysis presented above gets
modified by interaction between the eigenvalues which come
from the path integral measure, the result of which is that
that the eigenvalues behave like fermions.  As a result, 
the matrix model  is
described, in the double-scaling limit, by the second-quantized
fermion action \cite{SW,gross-kleb} (see \eq{a9})   
\be
\label{fermi-action}
H = \frac1{g_s}\int dx\ \Psi^\dagger(x)
[\frac{\hat p^2 - \hat q^2}{2} +1]\Psi(x), \quad [\hat q, \hat p]= i g_s
\ee
or by the bosonic variable $u(p,q,t)$ whose dynamics
is given by the classical action (see \eq{a:u-action})
\bea
\label{u-action}
S &=& \int\! dt\,ds\ \frac{dp dq}{2\pi g_s}
u (\del_t u \star \del_s u - \del_s u \star \del_t u)  
- \int\! dt\frac{dp dq}{2\pi g_s} 
u(p,q,t)(h(p,q) + 1) \nn 
&& h(p,q) = \frac{p^2-q^2}2
\eea
and the constraints
\bea
\label{constraints}
u\star u &=& u,
\nn 
\int \frac{dp dq}{2\pi g_s} u(p,q) &=& N 
\label{rank}
\eea
In the context of the double-scaled theory the last equation
is interpreted appropriately in the limit $N\to \infty$ (see below,
Eqns. \eq{rank-ripple},\eq{rank-d}, \eq{rank-n}).

For our purposes here, the action will not play a role other than
to yield  the equation of motion
\be
\label{u-eom}
\Big[\del_t  + (p\del_q + q\del_p)\Big] u(p,q,t)=0
\ee
which follows from the variation of the action \eq{u-action}
(see Appendix A). 

The appearance of the star product (see \eq{def-star}) indicates that
the field theory of $u(p,q,t)$ is noncommutative, reflecting the
noncommutative structure of the $p,q$ plane. The noncommutativity
parameter is the string coupling $g_s$. We will see below that it is
essentially the noncommutative nature of this bosonic theory that
prevents the divergence associated with the rolling tachyon.

\gap3

\noindent\underbar{Remarks on noncommutative solitons}:

\gap3

\noindent The equation $u\star u=u$ has reappeared in the context of
noncommutative solitons \cite{GMS}. The projector solution
\eq{projector} has also been rediscovered in that context.  In the
light of this development, the Fermi sea and the D-brane solution that
we will describe below can be identified as rank $N$ time independent
and time dependent solitons of $c=1$ theory (respectively).

\subsection{The solution}

We will now describe the solution of the above field theory that
describes the rolling tachyon on the unstable D0-brane. First
some preliminaries.

It is easy to solve the equation of motion \eq{u-eom} 
\be 
u(p,q,t)= u_{initial}(\bar p(t), \bar q(t))
\label{sol}
\ee
where
\[(\bar p(t), \bar
q(t)) = (p \cosh t - q \sinh t, - p \sinh t + q \cosh t)
\]
Note that time evolution preserves the area in phase space.
Since the constraints are preserved by the equation of motion (easy to
check) the simplest method of finding solutions to the equation of
motion as well as the constraints is to construct 
$u_{initial}(p,q,0)$ satisfying the constraints and use \eq{sol}.

We will construct various solutions by using the following observation
\cite{DMW-main,DMW-instanton}: the constraints simply mean the rank N
projector condition (see Appendix A). Thus, we should first construct
various rank $N$ projection operators $\hat u$ in the single-particle
Hilbert space and then convert it to $u(p,q)$ using
\eq{moyal}.

We begin with the solution corresponding to the Fermi sea.

\tgap

{\bf Fermi Sea:} 
\tgap
\[
\hat u= P_N = \sum_1^N \ket{\chi_\nu} \bra{\chi_\nu} \equiv \hat u_0
\]
It is straightforward to write down the corresponding 
function $u_0(p,q) = Tr(\hat u_0\ w(p,q))$ (see \eq{moyal}).

\tgap
{\bf Small fluctuations around the Fermi level and closed strings:}
\tgap

The small fluctuations or density waves around the Fermi level
(energies small compared with $g_s$) are described by an effective
boson theory that is described in Appendix C (Eqn.
\eq{collective}). This boson field $\phi$ is related to the closed
string massless mode (tachyon) by the well known leg pole transform
\cite{polchinski,natsuume,kutasov}. Since this sort of mapping is tied
to the $p^\pm$ parameterization of the classical Fermi fluid profile
which does not always work, in \cite{DMW-beta,DMW-discrete} the closed
string tachyon was mapped directly to a low energy fluctuation of the
phase space density $\delta u(p,q,t) = u(p,q,t) - u_0(p,q)$:
\bea
T(x,t) && = \int \frac{dp dq}{2\pi g_s} G_1(x;p,q)\delta u(p,q,t)
\nn
&& + \frac12 \int \frac{dp dq}{2\pi g_s}\int \frac{dp' dq'}{2\pi g_s}
G_2(x;p,q;p',q')\ \delta u(p,q,t)\ \delta u(p',q',t)
\nn
&& + ...
\label{full-transform}
\eea
The precise forms of $G_1, G_2$ are given in \cite{DMW-discrete}.
As the fermion fluctuation moves far to the left (away from
the turning point), the transform looks like
\be
T(x,t)=   \int \frac{dp\ dq}{2\pi g_s} f(-q e^{-x}/\sqrt{g_s})
\ \delta u(p,q,t) + O(x e^{-2x})
\label{transform}
\ee
where the function $f$ is given by a Bessel function
\[
f(\sigma) = \frac1{2\sqrt \pi} J_0\Bigg(2 
\Big(2/\pi\Big)^{1/8} 
\sqrt\sigma\Bigg)
\]
Far away from the turning point, a matrix model fluctuation 
around $q$ corresponds to a tachyon fluctuation roughly around 
$x \sim \ln (-q)$ but with a tail given by the above
equation. The precise relation between matrix model
fluctuations and tachyon fluctuations is both non-local and
non-linear, as seen above.

The equation of motion of the tachyon and its interactions can be
derived from the $u(p,q,t)$ dynamics \cite{DMW-beta,DMW-discrete}.It
is in this process that we see the emergence of closed string
backgrounds. In the present case the background is flat
2-dim. Minkowski spacetime and a linear dilaton. Departures from flat
spacetime begin to appear (in the symmetric matrix model) depending
upon how the Fermi sea is filled. In \cite{DMW-discrete} an unequal
filling of the Fermi sea on the two sides of the potential gave rise
to a curved spacetime in 2-dims.which corresponds to the asymptotic
form of the metric of the 2-dim. black hole
\cite{Mandal:1991tz,Elitzur:cb, Witten:1991yr}.

We should make a comment about the constraint \eq{rank}.
Since both $u_0$ and $u$ satisfy this constraint, 
\eq{rank} should be understood for the small
fluctuations  as 
\be
\int \frac{dp\ dq}{2\pi g_s} \delta u(p,q)= 0 
\label{rank-ripple}
\ee

\noindent{\bf D-brane:}
\tgap

\noindent 
We wish to describe a classical solution $u(p,q,t)$ which represents a
localized fermion high above the Fermi sea (Fig. 2). Since the rank of
$\hat u$ is always $N$ (cf. \eq{rank}), we must lift a fermion from
the Fermi sea and put it up. The first guess would be to put the
fermion up in an energy eigenstate $\psi_\nu$ (of energy $\nu$ far
above the Fermi sea)
\be
\label{try}
\hat u= \hat P_{N-1} + \ket{ \psi_\nu} \bra{ \psi_\nu}
\ee
However, it is easy to see that energy {\em eigenstates} are not
well-localized \cite{moore}. Thus for this fermion to be localized,
it must have a wave-function $\psi(q,t)$ which is a linear combination
of energy eigenstates. We will suppose that the wavefunction
$\ket{\psi(x,t)}$ is such that the phase space location of the fermion is
localized, within a size $\hbar = g_s$, around the point 
$(q_0, p_0)$. Such a wavefunction is given by (at $t=0$)
\[
\psi(x,0)= \exp\left[- \frac1{2g_s}\left((x- q_0)^2 + 2i p_0 x\right)
\right]
\]
The projector $\ket \psi \bra \psi $ is not orthogonal
to $\hat P_{N-1}$, however, since $\ket{\psi}$ is a linear
combination of an infinite number of energy eigenstates,
including those inside the Fermi sea! The naive solution
\eq{try} wouldn't satisfy $\hat u^2 = \hat u$, therefore.

The modification required is not difficult. We need to
project out from $\ket{\psi(x,0)}$ the components along
$P_{N-1}$ as in Gram-Schmidt orthogonalization. The result is

\[
u(p,q,t) = P_{N-1}(p,q) + u_1(p,q,t) - u_{01}(p,q,t)
\]
where
\be
u_1 = \exp[- \frac1{2g_s} \left((\bar q(t) - q_0)^2 
+ (\bar p(t) - p_0)^2\right)]
\label{u1}
\ee
and
\[\hat u_{01} = \hat 
P_{N-1}\ket{\psi_1}\bra{\psi_1} + \ket{\psi_1}\bra{\psi_1} 
\hat P_{N-1}
\]
We have used \eq{moyal} to switch back and forth between
operators and functions on phase space.
It is easy to check that (App. B) $u_{01}$ is negligible.
We will therefore write our solution as
\be
\label{full-u}
u(p,q,t) = u_0(p,q) + u_1(p,q,t)
\ee 
We have written $u_0(p,q)$ for $P_{N-1}(p,q)$ since in the
double scaling limit they correspond to the same function
(although the area is depleted by one, see \eq{rank-d}).
$u_1$ represents a localized wave packet in phase space which
is far from the D-brane at $t=0$, representing a D-brane.
The energy of $u_1$ is clearly 
\be
E_1 = \frac1{g_s}\Bigg(\frac{p_0^2-q_0^2}2 + 1\Bigg)
\label{e1}
\ee
where the quantity in the parenthesis is  of order one
by choice. 

It is clear from the above discussion that the solution
$u_1(p,q,t)$ satisfies the constraint
\be
\int \frac{dp\ dq}{2 \pi g_s} u_1(p,q,t) = 1
\label{rank-d}
\ee
which is different from \eq{rank-ripple} satisfied by
the small fluctuations (the reason is that the the background
$u_0$ here is one-fermion depleted).
For $n$ D0 branes see subsection \ref{sec:multi-d}.

It is useful to look at the position space density corresponding to
\eq{full-u}. We  have
\[
\rho(q,t)= \int \frac{dp}{2\pi g_s} u(p,q,t)
= \rho_0 + \rho_1
\]
where
$\rho(q) \to (1/g_s) \sqrt{q^2 -1}$ as $g_s \to 0$
represents the fermion density in the sea, and
\[
\rho_1(q,t) = \frac{1}{\sqrt{2\pi \Delta(t)}} \exp[-
\frac{(q - \bar q_0(t))^2}{ 2 \Delta(t)}]
\]
where
\[
\Delta(t) = \frac{g_s}2 \cosh 2t, \; \bar q_0(t)
= q_0 \cosh t + p_0 \sinh t
\]
represent, respectively, the time-dependent dispersion and the
trajectory of the centre of the wave-packet.

The time $T_c$ which needs to be crossed to reach the weak coupling
region, i.e. $|\bar q_0(t)| \gg o(1)$, gives a characteristic time scale
of the solution. Clearly
\[
T_c = O(1)
\]
(there is another time scale $t_0$ characterizing spreading of
the wave packet $\Delta(t_0) \sim o(1)$; $t_0 = -\ln \sqrt{g_s}$).

$\bullet$ For $t \ll T_c$, the solution
\eq{full-u} represents the Fermi sea $+$ 1 D0-brane.

It is easy to see (App. C) that at early times such as these the
solution $u_1(p,q,t)$ does not satisfy the equations which describe
the small fluctuations around the Fermi surface (effective coupling
$g_s/q$ must be small for such fluctuations). Alternatively, in this
region the perturbative definition \eq{full-transform} or
\eq{transform} breaks down.

$\bullet$  For $t \gg T_c$, by definition
$|\bar q_0(t)| \gg 1$, hence  the effective coupling
is small. In terms of a  phase space picture, note that
\[
u_1(p,q,t) \sim  \exp[- 
\left((p-q - p_0 e^{-t})^2 + (p - q + q_0
e^{-t})^2\right)/(8 g_s e^{-2t})]
\]
This describes a phase space density which is exponentially close to
the asymptote $p=q$, and is also close to the Fermi level $p =
\sqrt{q^2 - \mu} \sim q$; this means that the phase space density can
be represented as a small fluctuation of the Fermi surface. However,
this argument ignores the delocalization of the $p+q$ variable. A more
rigorous argument (see e.g.\cite{SW, MSW}) is to note that
non-relativistic corrections to energy levels $\epsilon$ above the
Fermi surface are controlled by the norm of the wavefunction
corresponding to the level $\epsilon$. Therefore the corrections to
the relativistic wavefunction go as $o(\epsilon/|q_0|^2)$ and become
$\ll 1$ for $t \gg T_c$. Hence in this regime the fermion wave-packet
can be decomposed into relativistic wave-functions.

It is possible to represent $u_1(p,q,t)$ at such late times as closed
string tachyon fluctuations using equations \eq{full-transform},
\eq{transform}.  Equivalently, it can be shown that the solution we
have found, at late times satisfies the equations of motion of density
waves near the Fermi surface.  The process of time-evolution of
$u_1(p,q,t)$ can, therefore, be interpreted as decay of a D-brane into
closed string modes.

\subsection{Absence of divergence}

Note that the solution that we have described here is {\bf classical},
still the description of the decay is free of divergence, since (as
\cite{KMS} have already argued and is clear from above) a fermion with
finitely localized wavefunction has a finite $\langle E_{total}
\rangle \sim 1/g_s$.  Such a description was not possible in the
standard {\rm classical} descriptions like BCFT, since the fermions
had sharply localized position {\em and} momenta. Here, however, we
have a classical (noncommutative) description, which is not limited in
such a fashion. Indeed, it can describe a finite fuzz of the particle
phase point because of the noncommutative nature of the classical
field theory.

Our formulation here also shows how to understand the conversion of
the D0 brane to tachyons by using \eq{full-transform}. Using this
equation (and its Fourier transform in the $x$-variable), we get the
distribution of tachyon quanta at various energies at late times (it
becomes a non-uniform distribution with finite total energy). The
total energy indeed turns out to be $1/g_s$.  The fact that at
finite $g_s$ the total energy of the tachyons is finite, and equal to
\eq{e1}, simply follows from the property that the Hamiltonian of the
tachyons is the same as the Hamiltonian of the $u_1(p,q,t)$ variable
(the transformation equations respect this fact). Since the energy of
$u_1(p,q,t)$ is conserved, it is always given by \eq{e1}, at early as
well as late times. At late times the map to tachyons is available,
hence the energy of the tachyon fluctuations can be equated to the
energy of $u_1$, namely \eq{e1}.

\subsection{More general forms of $u_1$}

It is not difficult to see that the specific Gaussian form of the
phase space density that we assumed above have not played any role for
our purposes here. The main point is that any localized wave packet
must satisfy the uncertainty relation $\Delta p \Delta q \sim
g_s$. All our conclusions can be shown to follow from this property.
However, the shape of the phase space density captures information
about the quantum properties of the D0-brane beyond the classical
property of position and velocity ($q_0, p_0$).

\subsection{Multiple D0 branes}\label{sec:multi-d}

It is easy to generalize the above discussion to
construct multiple D0 branes. Sticking to the Gaussian form for
simplicity, the solution is given by 
\bea
u(p,q,t) &&= u_0(p,q) + \delta u(p,q,t)
\nn
\delta u(p,q,t) &&= \sum_{j=1}^{n}  
 \exp[- \frac1{2g_s} \left((\bar q(t) - q_j)^2 
+ (\bar p(t) - p_j)^2\right)]
\label{multi-d}
\eea
Here $u_0$ represents the Fermi sea depleted by $n$ fermions
from the top (in the double-scaled limit it coincides with
the original Fermi sea). The centres of the Gaussians 
are chosen such that each $p_i, q_i \sim o(1)$ and for
each $i \not= j$ 
\be
(p_i - p_j)^2 + (q_i - q_j)^2 \gg o(g_s)
\label{centres}
\ee

As before, it is trivial to see that \eq{multi-d} 
satisfies the equation of motion. The condition \eq{rank} in this
case is
\be
\int \frac{dp\ dq}{2\pi g_s} \delta u(p,q,t)=n
\label{rank-n}
\ee
which is also easy to check. 

The constraint \eq{star} is more nontrivial. First, as in the case of
\eq{full-u}, one needs to show that the overlap of each of these D0
branes with the Fermi sea is small; this follows in a manner
similar to Appendix B. In addition, one needs to show that the
the overlap $u_{ij}$ between each pair $i,j$ of D-branes is small; it
follows that
\[
u_{ij} \sim \exp[- \frac{(p_i - p_j)^2 + (q_i - q_j)^2}{ 2 g_s}]
\]
This is small when the centres of the Gaussians are chosen
as in \eq{centres}.

\section{Conclusion}

\begin{quote}

$\bullet$ We have constructed a solution of 2-dim string theory valid
for arbitrary times. Before a characteristic time $T_c = o(1)$ it
describes a D-brane (plus Fermi sea). Later, it describes ripples
which can be translated into tachyon modes by appropriate integral
transforms. This constitutes a classical description of the rolling
tachyon which decays into closed string modes.

$\bullet$ The previous classical descriptions such as BCFT suffered
from divergences because the phase space location was infinitely
sharply localized in these classical descriptions. In the description
presented here, the field theory is noncommutative (with
noncommutativity parameter $g_s$). This allows for classical
solutions with fuzzy initial conditions (fuzzy phase space locations),
thereby leading to a finite result $\langle E_{total} \rangle= 1/g_s$.
It is an essential feature of noncommutative field theory to incorporate 
$g_s$ effects at the 'classical level'. 

\end{quote}

\gap4

\noindent{\bf Acknowledgments}

\gap2

\noindent 
We would like to to acknowledge useful discussions with Allan Adams,
Avinash Dhar, Rajesh Gopakumar, Antal Jevicki, Shiraz Minwalla and
Ashoke Sen.

\newpage

\appendix

\section{Review of string field theory of c=1}

We review salient features of the c=1 bosonic string field theory
developed in \cite{SW,DMW-main,DMW-beta,DMW-instanton}.

\begin{itemize}

\item  Non-interacting  fermions:
The partition function of the $c=1$ reads:
\[ Z= \int DM\ \exp[- S], \quad
S=  \frac1{g_0} \int dt\ Tr[ \frac12 (\dot M)^2 +  V(M)] \bigg), \qquad
\]
$V(M)= - M^2/2 - M^3$. The difference from \eq{action} is the absence
of the gauge field $A_0$. The inclusion of the gauge field by
\cite{KMS} amounts to restricting to the singlet sector of the above
partition function.  In this sector the theory reduces to that of $N$
eigenvalues of the matrix $M$. The result of integration over the
angles is that the eigenvalues $q_i$ behave as non-interacting
fermions. Each fermion is subject to a single-particle Hamiltonian
\be
h = g_0 \frac{p^2}2 + \frac1{g_0}( - \frac{q^2}2 - q^3)
\label{single-h}
\ee

\item Double-scaling: The single particle energy levels (eigenvalues
of $h$), ignoring tunneling out of the well, are as shown in
Fig. \ref{phi3}. The ground state of the matrix model is represented
by the N-fermion state in which the first $N$ levels are filled. The
location of the Fermi level, called $-|\mu_N|$, depends on
$g_0$. There is clearly a critical value $g_c$ of $g_0$ at which
$|\mu_N|\to 0$ i.e. the Fermi level reaches the maximum of the
potential, signaling a singularity of the partition function
$Z(g_0)$. The limit \be g_0 - g_c \approx \frac1{2\pi} \mu_N \ln \mu_N
\to 0, \hbox{or}, \mu_B \to 0,
\label{continuum}
\ee defines, therefore, a continuum limit of the random triangulation
represented by the matrix model. Consider, on the other hand, the 't
Hooft planar limit of the matrix model $N \to \infty, g_0 N = \bar g_0$
constant, in which only planar diagrams survive (genus zero). Double
scaling is defined in which these two limits are taken together: 
\be 
N \to \infty, \mu_N \to 0~ ({\rm or}~ g_0 \to g_c) \hbox{such that} \mu =
N \mu_N = \hbox{constant}
\label{double-scaling}
\ee
By setting up a WKB expansion of the wavefunctions of \eq{single-h}
(which gets arranged in powers of $1/\mu^2$), and identifying it
with the genus expansion of string theory, it 
can be seen that 
\be
\mu = 1/ g_s
\label{string-coupling}
\ee where $g_s$ is the string coupling.  

In the double scaling limit, the fermions are described in terms by a
scaled Hamiltonian written using second-quantized fermions 
\be H =
\int dq\ \Psi^\dagger(q)[h(\hat q, \hat p) + |\mu| ]\Psi(q)
\label{a:fermi-action}
\ee 
Here we have incorporated the information about the Fermi level
using $|\mu|$ as a Lagrange multiplier. The single particle
Hamiltonian becomes quadratic (the cubic term
scales away to zero)
\be
h = \frac12 (\hat p^2  - \hat q^2), \quad [\hat q,\hat p] =i   
\label{quadratic-h}
\ee

\underbar{Rescaling}:

To understand the semiclassical limit (cf. \eq{string-coupling}) it is
appropriate to perform a rescaling 
\be 
\hat q \to \sqrt g_s \hat
q, \hat p \to \sqrt g_s \hat p
\label{rescale}
\ee
so that 
\be
[\hat q,\hat p] =i g_s
\ee
This indicates that in the limit of small $g_s$, the 
one-particle phase space can be thought of as cells
of size $g_s$.

In this notation, the Hamiltonian becomes
\be
H = \frac1{g_s}\int dx\ \Psi^\dagger(x)[ h +1]\Psi(x)
\label{a9}
\ee
with $h$ still given by \eq{quadratic-h}.

We will denote energy levels of $h$ by \be
\label{levels}
h \chi_{\nu}(x) = \nu \chi_{\nu}(x)
\ee 
where $\chi_{\nu}(x)$ span the single particle 
Hilbert space ${\cal H}_1$.
The Fermi level is defined by the wavefunction
$\chi_{\nu}(x)$ with energy 
\be
\label{fermi-level}
\nu= - 1 \ee.

\item {\bf Construction of the bosonic field theory}


\myitem{\bf Example of  finite number of single-particle levels:}

Let us consider $N$ non-interacting fermions each of which can occupy
$K$ levels. Note that \eq{levels} have infinite number of levels, but
for simplicity we will consider first the case of $K,N$ finite.  The
limit $K,N \to \infty$ will be taken afterwards. The states of a
$K$-level system with $N$ levels filled can be described in terms of
the following overcomplete basis (coherent states): \be
\label{coherent}
\ket{\Psi_g} = \hat g \ket{F_N}
\ee
Here $\ket{F_N}$ is the filled Fermi sea, $g$ is a $U(K)$ group
element $\exp[i\theta_{\mu\nu} T_{\mu\nu}]$ and $\hat g$ is its
representation in the many-fermion system 
$\exp[i\theta_{\mu\nu} J_{\mu\nu}]$. Here
\be
\label{algebra}
\hat J_{\mu_\nu} = c^{\dagger}_{\mu}c_{\nu}
\ee
form a representation of $U(K)$ in the fermion
Fock space. $c_\nu, c^{\dagger}_\nu$ annihilates or creates (resp.)
the a fermion in the single-particle state $\chi_{\nu}(x)$.

It is clear that a path integral over the many-fermion system can be
converted to an integral over the group elements $g$, where $g$ varies
over the coset $U(K)/H$ where the subgroup $H=U(N) \times U(K-N)$
reflects the invariance group of the filled Fermi sea in
\eq{coherent}. Thus a (bosonic) description of the classical
configuration space can be provided by such group variables (see
spin-half example below). However, an alternative bosonic description
is provided by the variables
\be
\label{def-u}
u_{\mu\nu} = \mat{\Psi}{\hat g J_{\mu\nu}{\hat g}^\dagger}{\Psi}
= (g \bar u g^\dagger)_{\mu\nu}
\ee
with 
\be
\bar u_{\mu\nu} = \mat{F_N}{J_{\mu\nu}}{F_N}.
\label{u-base}
\ee
To elucidate, let us briefly consider the example of a spin-half
particle.

\noindent\underbar{Case of spin half particle}:
 
Let us consider the example of K=2, N=1, i.e. a two-level system with
half-filling. Two orthogonal basis states are $\ket{F_1} \equiv
c_1^\dagger \ket{0}$ and $ c_2^\dagger \ket{0}= c_2^\dagger
c_1\ket{F_1}$.  Here $c_1^\dagger, (c_2^\dagger)$ create a particle in
the state 1 (2 resp.); $\ket{0}$ is the no-particle state. These form
a spin-1/2, charge 1, representation of U(2)= SU(2) $\times $ U(1)
under the representation \eq{algebra}. The coherent state
\eq{coherent} in this case can be identified with the unit vector
${\mathbf n}$ in $R^3$ obtained by applying the rotation $g$ on a
given unit vector, say $\bar {\mathbf  n}= (0,0,1)$. The set of vectors
${\mathbf n}$ parameterize $U(2)/(U(1) \times U(1))=S^2$, the
classical configuration space of a spinning particle. The quantity
$u_{\mu\nu}$  in \eq{def-u} is simply related to this
spin variable:
\[
u_{\mu\nu} = {\mathbf n}.{\vec \sigma}_{\mu\nu} + \delta_{\mu\nu}
\]
To see this,  use the formula:
\[
\mat {\mathbf n}{\mathbf J}{\mathbf n} = -j {\mathbf n}
\]
where in our case $j=1/2$. The  base value $\bar u$ in\eq{def-u}
corresponds to the representative value of ${\mathbf n}$ in the
orbit. Here $\vec \sigma= \{\sigma_1, \sigma_2, \sigma_3\}$
are the Pauli matrices.

A spin-half particle moving in a magnetic field ${\mathbf B}$
translates to the following fermion equation:
\[
\del_t c^\mu(t)= {\mathbf B}.{\vec  \sigma}_{\mu\nu} c^\nu(t)
\]
The bosonic EOM is more familiar:
\[
\del_t {\mathbf n} = {\mathbf n}\times {\mathbf B} 
\]
obtainable from an action 
\be
\label{action-spin1/2}
S=\int dt\ ds\ Tr 
[ u(\del_t u \del_s u - \del_s u \del_t u) + B u]
\ee
where $B= {\mathbf B}.{\mathbf \sigma}$. Here $u(t,s)$ is a
one-dimensional extension of the classical variable $u(t)$.  It is
easy to visualize in terms of the equivalent variable ${\mathbf
n}(t)$; if ${\mathbf n}(t)$ traces a closed path in the configuration
space $S2$, then ${\mathbf n}(t,s)$ describes the solid angle enclosed
by the closed path. An infinitesimal variation of ${\mathbf n}(t,s)$
(which is an infinitisimal $SU(2)$ transformation) involves only the
boundary curve ${\mathbf n}(t)$, giving rise to the equation of motion
for the original variable ${\mathbf n(t)}$ written above.

Let us come back to the case of  general $K,N$. We now
understand  the $u$-variables as a classical spin variable,
characterizing a $G$-(coadjoint)-orbit,  ($G=U(K)$), of the
representative value \eq{u-base}. Using \eq{algebra},
the latter evaluates to
\be
\label{u-base-val}
\bar u_{\mu\nu} = \sum_{\lambda=1}^N
\delta_{\lambda\mu} \delta_{\lambda\nu}
= \left( \begin{array}{ccccccc}
1&&&&&&\\
&1&&&&&\\
&&.&&&&\\
&&&1&&&\\
&&&&0&&\\
&&&&&.&\\
&&&&&&0\end{array}
\right) 
\ee
where the first $N$ diagonal entries are 1 and rest are 0.
It is useful to regard the matrix $u_{\mu\nu}$ as an operator
in the first quantized Hilbert space ${\cal H}_1$: thus
\be
\label{operator}
u_{\mu\nu} =: \mat{\mu} {u} {\nu}, \ 
\bar u_{\mu\nu} =: \mat{\mu} {\bar u} {\nu}
\ee
Equation \eq{u-base-val} then implies that
$\bar u$ is a rank $N$ projector (onto  the
first $N$ single-particle levels)
\be
\bar u = P_N = \sum_{\lambda=1}^N \ket{\lambda}\bra{\lambda}
\ee
This implies
\be
\label{projector}
u = g P_N g^\dagger = \sum_{\lambda=1}^N \ket{\lambda_g}\bra{\lambda_g},
\quad  \ket{\lambda_g}\equiv g \ket{\lambda}
\ee
where the $U(K)$ matrix $g$ is interpreted to act on ${\cal H}_1$
in a manner similar to that in \eq{operator}. The common
property of the orbit \eq{projector} is that all $u$'s are
rank $N$ projectors, which is equivalent to the equations
\be
\label{a:constraints}
u^2 = u, \quad {\rm Tr}~ u= N
\ee

\myitem{\bf Limit $K\to \infty$ or $c=1$ model}

We now return to $c=1$.  Here the single-particle Hilbert space ${\cal
H}_1$ is infinite dimensional $K \to \infty$. The limiting case of
$U(K)$ is identified with the group of unitary operators $U({\cal
H}_1)$, and is called the group $W(\infty)$. The rest
of the analysis is pretty much unchanged. 

Operators $\hat u$ on the one-particle
Hilbert space ${\cal H}_1$ have an additional ``phase space''
representation (the Moyal map)
\bea
\label{moyal}
u(p,q)  &=& {\rm Tr}(\hat u \hat w(p,q))
\nn
\hat w(p,q) &=& \exp[\frac{i}{g_s}(q \hat p + p \hat q)]
\eea
The operators $\hat w(p,q)$ provide a basis of $W(\infty)$
(see \cite{DMW-main} where we have used the notation $\hat g(p,q)$
for $\hat w(p,q)$).

In terms of u(p,q), and switching to the rescaled phase space
coordinates \eq{rescale} the constraints \eq{constraints} become
\be
\label{star}
u\star u= u
\ee
\be
\int \frac{dp dq}{2\pi g_s} u(p,q)= N 
\label{a:rank}
\ee
The dynamics of the fermion
system can be written entirely in terms of the $u$-variables
\cite{DMW-main}(cf. \eq{action-spin1/2}):
\be
\label{a:u-action}
S= \int dt\ ds\ \frac{dp dq}{2\pi g_s}
 \Bigg(u(\del_t u \star \del_s u - \del_s u \star\del_t u ) + (h(p,q) + 1)
u(p,q)\Bigg)
\ee
where $u$ satisfies the two  constraint equations above. Here
$u(p,q,t,s)$ is an extension of the variable as described
below \eq{action-spin1/2}. Infinitesimal variation of
$u(p,q,t,s)$ involves only the ``boundary curve'' $u(p,q,t,s)$
(see discussion below \eq{action-spin1/2}), leading
to the equation of motion \eq{u-eom}. (Here the
variation of $u(p,q,t)$ is an infinitisimal $W(\infty)$
transformation, $\delta u= \epsilon \star u - u\star \epsilon $. This
variation also leaves the constraints \eq{constraints} invariant.)

This constitutes a {\bf noncommutative bosonic field theory} which
describes the c=1 matrix model. The star product, explicitly, turns
out to be 
\be (A\star B)(p,q) = \exp[i\frac{g_s}2(\del_q \del_{p'} - \del_p
\del_{q'})] (A(p,q) B(p',q'))|_{p=p',q=q'}
\label{def-star}
\ee 

\myitem{Role of fermions:}

The constraint \eq{star} simply reflects the projection operator
structure \eq{u-base-val}, which in turn reflects the fermionic
nature of the problem. The only non-zero diagonal entries are
1 because of Pauli exclusion principle.


\end{itemize}
\section{Computation of $u_{01}$}

The wavefunctions $\chi_{-\nu}(x)$ appearing in \eq{levels}
have the following asymptotic form \cite{Moore} relevant for
our purpose:
\be 
\nu \gg x^2: \quad 
\psi_\nu(x) \sim \nu^{-1/4} \exp[i x \sqrt \nu]
\ee
To get the Fermi level we put the condition \eq{fermi-level}.
Our wavefunction $\psi_1$ is given by 
\[
\psi_1(x)= \exp[- (x-q_0)^2/2 + i p_0 x]
\]
It is trivial to calculate the overlap:
\[
\langle\psi_1| \psi_\nu \rangle \propto 
\nu^{-1/4} \exp[- (\sqrt\nu + p_0)^2/2] \exp[i q_0 \sqrt \nu]
\]
For $\nu= \mu \sim 1/g$, and  $p_0, q_0 \sim O(1)$ this goes
as
\[ \exp[- 1/g_s] \]

\section{$u_1$ is not describable by collective field at early times}

The collective field limit:

In the limit $g_s \to 0$ the equation $u \star u=u$ becomes $u^2 =u$,
which means that $u(p,q,t)$ is 1 (in the region occupied by the
fermions) or else zero.  Such $u(p,q,t)$ are in simple cases
described by two $p$-extremities $p_+(q,t), p_-(q,t)$ of  
the region occupied by the fermions \cite{polchinski}.
In this case, the various moments of the phase space density: 
\bea m_0(q,t) 
&&=
\int \frac{dp}{2\pi g} u(p,q,t) = p_+ (q,t) - p_-(q,t) 
\nn  m_1(q,t) &&= \int \frac{dp}{2\pi
g}\ p\ u(p,q,t) = \frac12 ( p_+^2 (q,t) - p_-^2(q,t) ) 
\nn m_2(q,t) &&= \int \frac{dp}{2\pi g}\ p^2\
u(p,q,t) = \frac13 ( p_+^3 (q,t) - p_-^3(q,t) ) 
\nn ... && = ...
\label{moments}
\eea 
all get related to the two functions $p^\pm(q,t)$ and hence they
get related to the first two, which are
\[
m_0(q,t)= \rho(q,t), \; m_1(q,t)= \Pi(q,t) \rho(q,t)
\]
This is the situation for $g_s \to 0$
and for fluctuations not far from the Fermi sea. In this case, the
dynamics of such fluctuations is given by the following effective
action (the collective coordinate action)
\cite{das-jevicki,SW,gross-kleb,polchinski}:
\be S= \int dt\ dx[
\del_+ \phi \del_- \phi - \frac\pi{6 g_s} \frac1{\sinh2 x} \{(
\del_+ \phi )^3 -( \del_- \phi )^3 \}]
\label{collective}
\ee
where  $\del_\pm \phi$ are related to $\rho, \Pi$. 

It is easy to see that moments of $u_1(p,q,t)$ do not satisfy the
equation of motion that follow from \eq{collective} for $t \ll g_s$
\cite{DMW-instanton}. The extra terms leading to the
disagreement persist at large $t$. They are small but nonzero.

\end{document}